\documentclass[12pt]{article}
\usepackage{amsmath, amssymb}
\newcommand{\la}{\label}

\newcommand{\be}{\begin{equation}}
\newcommand{\ee}{\end{equation}}
\newcommand{\ba}{\begin{eqnarray}}
\newcommand{\ea}{\end{eqnarray}}
\newcommand{\bastar}{\begin{eqnarray*}}
\newcommand{\eastar}{\end{eqnarray*}}

\def\2{{1\over 2}}
\def\Lie{{\rm Lie}}

\begin{document}
%
%
%
\hoffset 0.5cm
\voffset -0.7cm
\evensidemargin 0.0in

\oddsidemargin 0.0in
\topmargin -0.0in
\textwidth 6.2in
\textheight 8.2in
%
%
%

\begin{titlepage}
\begin{flushright}
FIAN/TD-18/04\\
ITEP/TH-81/04 \\
UUITP-01/05
\end{flushright}
$~$
\vskip 0.5cm
\begin{center}
{ \bf \Large \bf YANG-MILLS, COMPLEX STRUCTURES
\\ \vskip 0.3cm
                AND CHERN'S LAST THEOREM \\
\vskip 0.6cm
\tt \normalsize To The Memory of Shiing-Shen Chern 
}
\end{center}

\vskip 0.5cm
\begin{center}
{\bf Andrei Marshakov$^{\sharp }$} {\tt  and }  
{\bf Antti J. Niemi$^{\star}$
} \\

\vskip 0.3cm

{\it $^\sharp$ Theory Department, Lebedev Physics Institute\\
Leninsky Prospect 53, 119991 Moscow, Russia \\
and  \\
Institute of Theoretical and Experimental Physics \\
B.~Cheremushkinskaya 25, 117259 Moscow, Russia} 

\vskip 0.3cm
{\it $^\star$Department of Theoretical Physics,
Uppsala University\\ 
P.O. Box 803, S-75108  Uppsala, Sweden  \\
and \\
Nankai Institute of Mathematics, Nankai University \\
Tianjin 300071, P.R. China }\\
\end{center}
\vskip 1.cm

\rm
\noindent
Recently Shiing-Shen Chern suggested that the 
six dimensional sphere $\mathbb{S}^6$ has no complex 
structure. Here we explore the relations between his arguments
and Yang-Mills theories. In particular, we 
propose that Chern's approach is widely 
applicable to investigate connections between the 
geometry of manifolds and the structure 
of gauge theories. We also discuss several
examples of manifolds, both with and without 
a complex structure.
\noindent\vfill
\vskip 0.5cm
\begin{flushleft}
\rule{5.1 in}{.007 in} \\
{\small  E-mail: \scriptsize
\bf $^{\sharp}$ MARS@LPI.RU, MARS@ITEP.RU $\ \ 
\  \ $ $^{\star}$NIEMI@TEORFYS.UU.SE \\
\tt $^{\star}$http://www.teorfys.uu.se/people/antti/}
\\
\end{flushleft}
\end{titlepage}

%
%
\section{Introduction}
The existence of a complex structure has a central role
in many physical problems. It is of particular importance
to supersymmetric quantum field and string
theories, where geometry plays an essential role.
For example Ricci flat K\"ahler manifolds are 
fundamental constituents in superstring compactification 
\cite{gsw}, and mirror symmetry and 
duality relations involve structures that relate complex 
geometry to symplectic geometry \cite{polc}.
 
Sometimes it can be quite difficult to determine whether 
a manifold admits a complex structure. A notorious example 
in this respect is the six dimensional sphere $\mathbb{S}^6$.
The existence of a proof that it admits no complex structure
has been both widely advertised \cite{gsw}, \cite{naka} 
and openly challenged \cite{chern1}. Very recently 
Chern attempted to settle this controversy. At the age of
93 he outlined a proof that $\mathbb{S}^6$ has no complex 
structure \cite{chern2}. Sadly, he passed away before
presenting a complete proof of his Last Theorem.
Chern's approach employs differential geometry in a form
reminiscent of Yang-Mills theories. Thus we expect that 
generalizations of his approach will eventually lead 
to novel and deep connections between the geometry of 
manifolds and the structure of gauge theories. 

The scope of the present article is to broaden Chern's approach. 
We explain how it can be applied to study the 
properties of gauge theories and the existence of 
complex structures in a general class of manifolds.
As concrete examples we consider several familiar
gauge theory models with physical relevance. These 
describe the two dimensional sphere $\mathbb{S}^2$, 
where the existence of a complex structure is related to the 
structure of Dirac monopole. The four dimensional sphere 
$\mathbb{S}^4$, where the lack of a complex structure can be 
directly inferred from the properties of a Yang-Mills instanton. 
And the two (complex) dimensional projective space 
$\mathbb{CP}^2$ whose complex structure reflects 
the properties of the $SU(2) \times U(1)$ Lie subgroup 
in a $SU(3)$ Yang-Mills theory.

For a vector space that has $2n$ real dimensions,
a complex structure is a linear endomorphism
$\mathbb{J}$ that squares to $\mathbb{J}^2 = -1$. 
If the vector space is a tangent space of a 
$2n$-dimensional manifold $\cal M$, the set of  
endomorphisms $\mathbb{J}_x$ on the tangent bundle 
$T\! {\cal M}$ equips the manifold with an {\em almost} 
complex structure. The almost complex 
structure $\mathbb{J}_x$ is integrable and $\cal M$ 
is a complex manifold, iff \cite{chern1} 
\[
d = \partial + \bar\partial
\] 
The integrability can also be stated as
\[
\partial^2 \ = \ \bar\partial^2 \ = \ 0
\]
In analogy, on the cotangent bundle  $T^*\! {\cal M}$ 
of the manifold $\cal M$ we can
employ $\mathbb{J}_x$ to introduce globally defined 
$(1,0)$ one-forms $\omega_k$ ($k=1,\ldots,n$). 
If these one-forms on $T^*\! {\cal M}$ can be 
represented as holomorphic linear combinations of one-forms
$dw^k = du^k + idv^k$ where $(u^k,v^k)$ are local 
coordinates on $\cal M$, the almost complex structure 
$\mathbb{J}_x$ is integrable \cite{chern1}.

In a complementary description (see e.g. \cite{hitch}, \cite{gualt})
the integrability of an almost complex structure is formulated directly
in terms of the linearly independent type $(1,0)$ one-forms $\omega_k$
such that

\vskip 0.2cm

{\it i)} The almost complex structure on a manifold ${\cal M}$ is a
locally decomposable $(n,0)$-form 
$\Omega$,
\begin{equation} 
\Omega = \omega_1 \wedge \ldots \wedge \omega_n 
\la{hitc1}
\end{equation}

{\it ii)} $\Omega$ is non-degenerate,
\begin{equation}
\Omega \wedge \bar \Omega \ \not= \ 0
\la{hitc3}
\end{equation}

{\it iii)} The $(1,0)$-forms $\omega_k$ satisfy the 
integrability condition
\begin{equation}
\Omega\wedge d\omega_k\ =\ 0
\label{hitc2a}
\end{equation}
which can also be written in the alternative form
\begin{equation}
d\omega_k \ = \ 0 \ \ \ \ \ \ \ {\it modulo}\ (\omega_l)
\la{hitc2}
\end{equation}

From (\ref{hitc2}) it is obvious that {\it any} two-dimensional 
almost complex structure is integrable. A familiar example is the 
two dimensional sphere $\mathbb{S}^2\cong\mathbb{CP}^1$ which inherits 
its (unique) complex structure from the complex plane $\mathbb{C}$ by 
stereographic projection. But in higher dimensions 
a given almost complex structure on a manifold ${\cal M}$ is not 
necessarily integrable, and an almost complex manifold does not need 
to be a complex manifold. Examples with non-integrable
almost complex structures can be constructed whenever the real 
dimensionality $2n$ of a manifold is at least four 
\cite{chern1}. Furthermore, unlike $\mathbb{S}^2$ 
the four dimensional sphere $\mathbb{S}^4$ does 
not even admit an almost complex 
structure, a nontrivial fact that follows from 
the properties of its Chern class and index theorems.
Indeed, this lack of even an almost complex structure is a property
shared by {\em all} higher dimensional spheres 
$\mathbb{S}^{2n}$ whenever $n>3$. But the case $n=3$ is exceptional: 
The six dimensional sphere $\mathbb{S}^6$ is an almost complex 
manifold and it remains 
an open problem whether or not it is actually
a complex manifold.

\section{Chern's Last Theorem}

There is a good reason why $\mathbb{S}^2$ and $\mathbb{S}^6$ 
are the only
spheres that can have an almost complex structure. This is due
to the fact, that $\mathbb{R}^3$ and $\mathbb{R}^7$ are 
the only vector spaces where one can define an antisymmetric
bilinear cross product of vectors 
(see, e.g. \cite{baez}). 
This existence of a vector 
cross product on $\mathbb{R}^3$ and $\mathbb{R}^7$ 
reflects the fact that besides real and complex numbers, 
quaternions and octonions are the only normed division algebras. 

In the present section we consider the seven dimensional 
case $\mathbb{R}^7$,
where the vector cross product derives from the 
multiplicative properties of imaginary 
octonions \cite{baez}. 
Indeed, as a linear space imaginary octonions
coincide with $\mathbb{R}^7$. Thus the unit
sphere $\mathbb{S}^6 \subset \mathbb{R}^7$ is isomorphic with 
the space of unit imaginary octonions, and  
it acquires a natural almost complex structure 
from the action of the octonionic vector cross product 
in $\mathbb{R}^7$. Explicitely this octonionic 
almost complex structure on $\mathbb{S}^6$
is constructed as follows:
Let $\hat e_1, \dots , \hat e_7$ be the basis of imaginary unit 
octonions {\it i.e.} each of them is
a square root of $-1$. A generic point $y \in 
\mathbb{R}^7$ can be identified with 
\[
y = y_1 \hat e_1 + \ldots + y_7 \hat e_7
\]
A point $x$ on the unit sphere $\mathbb{S}^6 \subset 
\mathbb{R}^7$ becomes then identified with 
\[
x = x_1 \hat e_1 + \ldots + x_7 \hat e_7 \ ; \ \ \ \ \ \sum_i x_i^2 = 1
\] 
The octonionic multiplication of vectors in $\mathbb{R}^7$,
which we denote by $\times$, is defined in terms of the
octonionic multiplicative rule of the basis elements $\hat e_k$. 
This is a totally antisymmetric bilinear 
automorphism of the form
\begin{equation}
\hat e_i \times \hat e_j \ = \ \sum_{k=1}^{7} c^{ijk} \hat e_k
\end{equation}
We can choose the octonionic basis so that the only
nonvanishing components of the totally antisymmetric
octonionic tensor $c^{ijk}$ are \cite{murat}
\begin{equation}
c^{123} = c^{147} = c^{165} = c^{246} = 
c^{257} = c^{354} = c^{367} = 1
\end{equation}
This reflects the index cycling and the index doubling 
symmetries of the octonionic product,
\begin{equation}
\hat e_i \times \hat e_j = \hat e_k \ \ \Rightarrow \ \ \hat
e_{i+1} \times
\hat e_{j+1} = \hat e_{k+1} \ \ \ (mod \ 7) 
\end{equation}
\begin{equation}
\hat e_i \times \hat e_j = \hat e_k \ \ \Rightarrow \ \ \hat
e_{2i} \times
\hat e_{2j} = \hat e_{2k}
\end{equation}

Consider a $y \in \mathbb{R}^7$ which is orthogonal 
to a given $x \in \mathbb{S}^6 \subset \mathbb{R}^7$ in the sense of 
the standard $\mathbb{R}^7$ vector inner product; in the octonionic
basis we realize this inner product with
\[
<\hat e_i , \hat e_j> = \delta_{ij}
\]
The octonionic 
(cross) product $x \times y$ determines a mapping
\begin{equation}
y = y_i \hat e_i \ \to \ \mathbb{J}_x (y) \ 
\equiv x \times y \ = \ c^{ijk} x_i y_j \hat e_k
\la{J7vect}
\end{equation}
and due to the total antisymmetry of the $c^{ijk}$, we have
\[
\mathbb{J}_x(x) = 0
\]
The mapping (\ref{J7vect}) is clearly a linear automorphism 
$\mathbb{J}_x: \ y \to x \times y$ 
of the tangent bundle $T\mathbb{S}^6$. Explicitely, the action of
$\mathbb{J}_x$ on the tangent bundle coincides with that 
of the matrix
\begin{equation}
\mathbb{J}_x \ = \ \left(
\begin{array}{ccccccc}
0 & x_3 & -x_2 & -x_5 & x_4 & -x_7 & x_6 \\
-x_3 & 0 & x_1 & -x_6 & x_7 & x_4 & -x_5 \\
x_2 & -x_1 & 0 & -x_7 & -x_6 & x_5 & x_4 \\
x_5 & x_6 & x_7 & 0 & -x_1 & -x_2 & -x_3 \\
-x_4 & -x_7 & x_6 & x_1 & 0 & -x_3 & x_2 \\
x_7 & -x_4 & -x_5 & x_2 & x_3 & 0 & -x_1 \\
-x_6 & x_5 & -x_4 & x_3 & -x_2 & x_1 & 0 \\
\end{array}\right)
\label{matr}
\end{equation}
Since $|x| =1$ we clearly have
\begin{equation}
\label{J1m7}
\left(\mathbb{J}_x^2\right)_{ij} = 
- \delta_{ij} + x_ix_j
\end{equation}
This implies that for any tangent vector 
$y \in T\mathbb{S}_x^6$ 
\begin{equation}
(\mathbb{J}_x^2)_{ij}y_j \ = \ -y_i
\nonumber
\end{equation}
which confirms that $\mathbb{J}_x$ indeed defines an almost 
complex structure on $\mathbb{S}^6$.

The present (octonionic) almost complex structure on
$\mathbb{S}^6$ is {\em not} integrable. This  
can be verified by a direct computation of (\ref{hitc2}), 
for example with the one-forms 
\[
\eta_a = idx_a+\sum_{b\neq a} 
J_{ab}dx_b
\]
with 
\[
J =
{\mathbb{J}_x}_{
|_{x_7=\pm \sqrt{1-\sum_{a=1}^6 x_a^2}
}}
\]
But it is in 
principle possible that in addition to a non-integrable 
complex structure there can exist an integrable one. The
deformation theory of (almost) complex structures is
described by the Kodaira-Spencer theory which we note, is 
also relevant 
to the topological type-B string theory \cite{BCOV}.

Recently Chern proposed \cite{chern2} that {\em none} of the 
almost complex structures on $\mathbb{S}^6$ can be
integrable, and argued that from this it follows
that $\mathbb{S}^6$ does not admit any complex structure. 
His approach is based on the fact that the almost
complex structure (\ref{J7vect}) is invariant under the natural
action of the 14-dimensional exceptional 
Lie group $G_2$ on $\mathbb{S}^6$. Indeed, the group $G_2$ is the 
automorphism group of octonions and 
as a manifold it is a principal $SU(3)$-bundle 
over $\mathbb{S}^6$, 
i.e. locally
\[
G_2 \sim \mathbb{S}^6 \times SU(3)
\]

In order to outline Chern's approach, we
denote by $\mathfrak{g}_i$ ($i,j,k=1,\ldots,14$) the Lie algebra 
generators of $\Lie(G_2)$, and by $\mathfrak{h}_a$ ($a,b,c=1,\ldots,8$)
we denote the generators of its $su(3)=\Lie (SU(3))$ subalgebra. 
Let $\mathfrak{m}_s$ ($s,t,u = 1,\ldots,6$) be the 
remaining generators of $\Lie(G_2)$ 
so that as vector spaces 
\[
\mathfrak{g} = \mathfrak{h} + \mathfrak{m}
\]
Under this decomposition the commutation relations of the 
$G_2$ Lie algebra acquire the form \cite{murat}
\begin{equation}
[\mathfrak{h}_a , \mathfrak{h}_b ] \ = \ f^{abc} \mathfrak{h}_c
\la{li1}
\end{equation}
\begin{equation}
[ \mathfrak{h}_a, \mathfrak{m}_{s} ] \ = \  
t^{ast} \mathfrak{m}_{t}
\la{li2}
\end{equation}
\begin{equation}
[ \mathfrak{m}_s , \mathfrak{m}_t ] \ = 
t^{sta}\mathfrak{h}_a + k^{stu}\mathfrak{m}_u
\la{li3}
\end{equation}
where all structure constants are totally antisymmetric.
Explicitely, we have for the nonvanishing $k^{stu}$ \cite{murat}: 
\[
k^{136} =  - k^{145} =  k^{235} = k^{246} = -\frac{1}{\sqrt{3}}
\]
Since these are nontrivial, we conclude that $su(3)$ is not a 
symmetric subalgebra of the $\Lie(G_2)$ Lie algebra. 
It is this nontriviality of $k^{stu}$ that ultimately
leads to the lack of integrability in the (invariant) almost complex 
structures on the sphere $\mathbb{S}^6$.

The relevant geometry of the Lie group $G_2$ can be 
described by the ensuing Maurer-Cartan equation,
which we write as a flatness condition for a $\Lie(G_2)$-valued 
Yang-Mills connection \cite{chern2}, \cite{bryant}
\begin{equation}
F \ = \ dA + A\wedge A = 0 \ \ \Leftrightarrow \ \ A = g^{-1} d g
\la{mc}
\end{equation}
Using the fact that $G_2$ is a $SU(3)$-bundle over 
$\mathbb{S}^6$, we decompose the $\Lie(G_2)$-valued 
Yang-Mills connection $A$ into a linear combination 
\[
A = \kappa + \rho
\]
where $\kappa$ takes values in the $su(3)$ subalgebra
and $\rho$ is a linear combination of the remaining 
six generators $\mathfrak{m}_s$. 

In terms of $\kappa$ and $\rho$
the components of the Maurer-Cartan equation 
acquire the form
\begin{equation}
Fâ_\kappa \ = \ d \kappa^a +  
f^{abc} \kappa^b \wedge \kappa^c \ = \ - t^{ast}\rho^s
\wedge \rho^t
\la{mcar1}
\end{equation}
\begin{equation}
d\rho^s \ = \ - 2t^{sat} \kappa^a \wedge \rho^t - k^{stu} \rho^t
\wedge \rho^u
\la{mcar2}
\end{equation}
Here $F_\kappa$ is the $SU(3)$ curvature two-form, and
the one-forms $\rho^a$ form a basis for the cotangent 
bundle $T^*\mathbb{S}^6$.  Since
$\mathbb{S}^6$ is an almost complex manifold 
these Maurer-Cartan equations can be represented in a 
manifestly complex form.
For this we introduce a holomorphic polarization
on $T^* \mathbb{S}^6$ and present the $\rho^a$ as 
linear combinations of the ensuing $(1,0)$-forms $\theta^\alpha$ 
and $(0,1)$-forms $\bar\theta^\alpha$ 
where now $\alpha,\beta,\gamma = 1,2,3$. Explicitly, one can choose
\[
\theta^1 \ = \ \rho^1 + i \rho^2
\]
\[
\theta^2 \ = \ \rho^4 +i \rho^3 
\]
\[
\theta^3 \ = \ \rho^6 - i \rho^5
\]
for the basis of $\Lie(G_2)$ algebra described in \cite{murat}.
The Maurer-Cartan equation (\ref{mcar2}) now acquires
the form \cite{bryant}
\begin{equation}
d\theta^\alpha \ = \ 
t^{\alpha a \beta} \kappa^a \wedge \theta^\beta
\ + \ \frac{2}{\sqrt{3}}\ \epsilon^{\alpha\beta\gamma}
\bar\theta^\beta \wedge \bar \theta^\gamma
\la{cmc2}
\end{equation} 
where the almost complex structure of $\mathbb{S}^6 \subset G_2$ 
is manifest. But since the last term in (\ref{cmc2}) involves
only the $(0,1)$-forms $\bar\theta^\alpha$ we have
\[
\Omega \wedge d\theta^\alpha \ \equiv \ 
\theta^1 \wedge \theta^2 \wedge \theta^3 \wedge d\theta^\alpha 
\ \not= \ 0 \ \ \ \ \ (\alpha = 1,2,3)
\]
which establishes that the integrability condition (\ref{hitc2a}), 
(\ref{hitc2}) is not obeyed. As a consequence 
the present almost complex structure is not integrable.

In \cite{chern2} Chern suggests that the existence of the
almost complex structure determined by the present one-forms 
$\left(\theta^\alpha, \ \bar\theta^\alpha\right)\in 
T^* \mathbb{S}^6$ makes it impossible to endow $\mathbb{S}^6$
with an integrable almost complex structure.
For this, he assumes that there is a complete
set of one-forms $\left(\omega, \bar\omega\right)$ 
that defines an {\em integrable}
almost complex structure on $\mathbb{S}^6$. By completeness 
one can represent these one-forms as linear combinations 
of the present one-forms $\theta, \ \bar\theta$,
\begin{equation}
\omega^\alpha \ = \ {p^\alpha}_\beta \theta^\beta +  {q^\alpha}_\beta
\bar\theta^\beta
\label{gen}
\end{equation}
Using linear algebra Chern argues \cite{chern2}
that {\em any} possible choice of ${p^\alpha}_\beta $ 
and ${q^\alpha}_\beta$ in (\ref{gen}) leads to a 
conflict between the integrability condition
(\ref{hitc2}) and the Maurer-Cartan equations (\ref{cmc2})
for $\theta^\alpha, \ \bar\theta^\alpha$.
A completion of his arguments should prove that the six dimensional 
sphere $\mathbb{S}^6$ has {\em no} complex structure.

\section{On General Structure}
Chern's approach \cite{chern2}
to the almost complex structure on $\mathbb{S}^6$ 
employs natural structures in Yang-Mills theory.
Consequently we expect that his approach has a much wider applicability. 
To inspect this, we consider a Yang-Mills theory with a gauge group
$\cal H$ on a manifold $\cal M$ so that together these two combine 
into a total space which is another Lie group $\cal G$. 
A physical perspective to this structure could be, that we 
have a $\cal G$-invariant gauge theory where the total group 
contains the gauge part $\cal H$ and the part
corresponding to the vacuum or space-time manifold $\cal M$. 
In a sense, the manifolds $\cal M$ and $\cal H$ are then dual
to each other within the gauge group $\cal G$.
The duality relation is a mapping between these two manifolds
and it is determined by the flatness condition 
\begin{equation}
F = dA + A\wedge A \ = \ 0
\la{ymgen}
\end{equation}
for the $\Lie({\cal G})$-valued Yang-Mills connection $A$
  
If we write $A = g^{-1} d g$ where $g \in \cal G$, the flatness
condition (\ref{ymgen}) leads to the standard form of the
Maurer-Cartan equation for $\cal G$. But instead we now
decompose $A$ as follows,
\begin{equation}
A \ = \ g^{-1} dg \ = \ \kappa + \vartheta
\la{decom}
\end{equation}
Here $\kappa $ is the projection of $A$ to the Lie algebra 
of the subgroup $\cal H$ and $\vartheta$ denotes the remaining
"space-time" components of $A$ in $\Lie(\cal G)$. We decompose
the Lie algebra of $\cal G$ into the ensuing vector space sum 
\[
\mathfrak{g} = \mathfrak{h} + \mathfrak{m}
\]
so that $\kappa$ is a linear combination of generators in
$\mathfrak{h}$, while $\vartheta$ is a linear combination of 
generators in $\mathfrak{m}$.
Under this decomposition the Lie algebra of $\cal G$ in 
general acquires the form 
\begin{equation}
[\mathfrak{h}_a , \mathfrak{h}_b ] \ = \ f^{abc} \mathfrak{h}_c
\la{lie1}
\end{equation}
\begin{equation}
[ \mathfrak{h}_a, \mathfrak{m}_{s} ] \ = \  
C^{ast} \mathfrak{m}_{t}
\la{lie2}
\end{equation}
\begin{equation}
[ \mathfrak{m}_s , \mathfrak{m}_t ] \ = 
C^{sta}\mathfrak{h}_a + C^{stu}\mathfrak{m}_u
\la{lie3}
\end{equation}
With these, the flatness condition (\ref{ymgen})
(Maurer-Cartan equation) becomes
\begin{equation}
F^a_\kappa \ \equiv \
d \kappa^a \ + \ f^{abc}\kappa^b \wedge \kappa^c \ = 
\ - C^{ast}\vartheta^s \wedge \vartheta^t
\la{genmc1}
\end{equation}
\begin{equation}
D_\kappa^{st} \vartheta^t \ \equiv \ 
(\delta^{st} d + 2 C^{sat}\kappa^a  )\wedge \vartheta^t \ = \ 
- C^{stu}\vartheta^t \wedge \vartheta^u
\la{genmc2}
\end{equation}
which one can now interpret as a certain "duality relation" between 
the curvature tensor $F_\kappa$ of the $\cal H$ 
invariant Yang-Mills theory with base manifold $\cal M$
and a ``monopole 
equation'' for $\vartheta$ on $\cal M$, coupled to 
those components of $\kappa^a$ for which $C^{sat}\not=0$. 
We note that a comparison with the functional form of Cartan's first 
structure equation \cite{eguchi} (see also
equation (\ref{cart}) below) suggests that the 
$\vartheta\wedge \vartheta$ term on the {\it r.h.s.} 
of (\ref{genmc2})
is a contribution from a torsion connection, 
and we see that it is directly related to the 
non symmetric structure of the Lie algebra decomposition 
(\ref{lie3}).

One can verify that these equations are gauge covariant {\it w.r.t.}
gauge transformations in the $\cal H$ subgroup. Indeed, if 
$h \in \cal H$ the ensuing gauge transformation acts as follows,
\[
\kappa \to h^{-1} \kappa h + h^{-1} d h
\]
\[
F_\kappa \to h^{-1} F_\kappa h 
\]
\[
\vartheta \to h^{-1} \vartheta h
\]
and using this in (\ref{genmc1}), (\ref{genmc2}) we find
that the equations are covariant under the $\cal H$ gauge
transformations. 

The equations (\ref{genmc1}), (\ref{genmc2}) can 
be employed to investigate various properties of the $\cal H$ gauge
theory and the geometry of the manifold $\cal M$. Of particular
interest to us here is whether $\cal M$ admits an almost 
complex structure and whether this structure can be integrable. 
For this we recall that the $\vartheta$ span 
the cotangent bundle $T^* \cal M$. If $\cal M$ is 
an almost complex manifold, we can introduce
a holomorphic polarization on $T^*  \cal M$ 
to represent the $\vartheta$ as 
linear combinations of the one-forms which are either of type 
$(1,0)$ or type $(0,1)$. 
For an almost complex 
$\cal M$, the Maurer-Cartan equations (\ref{genmc1}), 
(\ref{genmc2}) should then acquire a complex decomposition.
On the other hand, 
if $\cal M$ has no almost complex structure the 
Maurer-Cartan equations do not admit any complex decomposition, in 
any holomorphic polarization on $T^* \cal M$.

Suppose now the manifold $\cal M$ admits an 
almost complex structure, and that $\theta^\alpha , 
\ \bar\theta^\alpha$ are
one-forms of the type $(1,0)$ and $(0,1)$ respectively in
the given holomorphic polarization on $T^* \cal M$.
In such a basis the Maurer-Cartan equation (\ref{genmc2}) 
acquires the manifestly complex form 
\[
(\delta^{\alpha \beta} d + C^{\alpha a \beta } \kappa^a \wedge) 
\hskip 0.5mm
\theta^\beta \ = \
- C_{++}^{\alpha \beta \gamma} \theta^\beta \wedge
\theta^\gamma - C_{+-}^{\alpha\beta\gamma} \theta^\beta \wedge
\bar\theta^\gamma - C_{--}^{\alpha\beta\gamma} \bar\theta^\beta 
\wedge \bar\theta^\gamma
\]
see also (\ref{genmc2}), where the 
$C$-coefficients on the {\em r.h.s.} 
are in general complex-valued
functions that display the following
antisymmetries {\it w.r.t.} last two indices
\[
C_{++}^{\alpha \beta \gamma} + C_{++}^{\alpha\gamma\beta} \ 
= \
C_{--}^{\alpha\beta\gamma} + C_{--}^{\alpha\gamma\beta} 
\ = \ 0
\]
From (\ref{hitc2}) we conclude that the integrability of the
almost complex structure becomes translated into
the condition
\[
C_{--}^{\alpha\beta\gamma} \ = \ 0
\]
Clearly this condition is invariant under linear 
transformations of the $\theta^\alpha$. As a consequence 
the lack of integrability 
in the almost complex structure becomes related to the 
nontriviality of the structure constant $C^{stu}$ in (\ref{lie3}), 
which we already noted in connection of $\mathbb{S}^6$; 
see (\ref{li3}). This suggests that if 
the embedding of $\cal H$ in $\cal G$ is not symmetric 
the almost complex structure of $\cal M$ can not be integrable.

\vskip 0.2cm
\noindent
We conclude this section with the following observations:
\vskip 0.2cm

- If $C^{ast}$ vanishes the $\cal H$-curvature $F_\kappa$
is flat, and $\kappa$ is the Maurer-Cartan form of $\cal H$.

- If $C^{ast}$ is nontrivial, the connection $\kappa$
is nontrivial. The duality relation (\ref{genmc1}) 
then allows us to relate the Chern classes of the
principal $\cal H$ bundle over $\cal M$ to the 
solutions of the ``monopole equation'' defined directly on 
$\cal M$, since according to (\ref{genmc1})
\begin{equation}
{\rm Det}\left( 1 + \frac{1}{4\pi} F^a\mathfrak{h}_a \right)
\ = \ {\rm Det} \left( 1 - \frac{1}{4\pi} \mathfrak{h}_a 
C^{ast} \vartheta^s
\wedge \vartheta^t \right)
\la{class}
\end{equation}

- If $C^{ast}$ and $C^{stu}$ both vanish $\vartheta$ is closed
and $F_\kappa$ is flat. 

- If $C^{ast}$ vanishes but $C^{stu}$
does not, $F_\kappa$ is flat and (\ref{genmc2}) becomes
\[
F_\vartheta = 
d \vartheta^s + C^{stu} \vartheta^t \wedge \vartheta^u \ = \ 0
\]
Hence $F_\vartheta$ is in general 
a curvature with a torsion connection, since in general 
the $\mathfrak{m}_s$ do not define a Lie algebra.

- Finally, for the symmetric case, when $C^{stu}$ vanishes but 
$C^{ast} \not = 0$, the 
curvature $F_\kappa$ is nontrivial and the $\vartheta$ are covariantly
constant with respect to $\cal H$ and the Maurer-Cartan 
equations have the form
\begin{equation}
F^a_\kappa \ = \ - C^{ast} \vartheta^s \wedge \vartheta^t
\la{ws1}
\end{equation}
\begin{equation}
D_\kappa \vartheta \ = \ 0
\la{ws2}
\end{equation}
where we remind that $D_\kappa$ is a covariant derivative with
respect to those components $\kappa^a$ for which $C^{sat}\not=0$.
The functional form of these equations resembles the
Seiberg-Witten equations \cite{wse1}, \cite{wse2}. 
In particular, if as in
\cite{fad} we expand $\vartheta$ in terms of 
a hyperplane this resemblance becomes manifest. Indeed,
the duality considered here generalize 
the dualities considered in \cite{fad}.

\section{Two Dimensional Sphere}

It is a well known fact, that in three dimensions the familiar 
cyclic cross product of the orthonormal basis vectors 
$({\bf \hat i} , {\bf \hat j} , {\bf \hat k})\in\mathbb{R}^3$ 
coincides with the multiplicative structure of imaginary 
quaternions. Furthermore, as the 
sphere $\mathbb{S}^2 \subset \mathbb{R}^3$ 
is in a one-to-one correspondence with three dimensional 
unit vectors,
its (almost) complex structure is a manifestation 
of the multiplicative properties of unit imaginary quaternions.
We now consider this case as an illustrative but simple
example of the previous constructions.

We introduce 
a $U(1) \simeq \mathbb{S}^1$ gauge 
theory on $\mathbb{S}^2$. 
Since locally $\mathbb{S}^2 \times \mathbb{S}^1 \simeq 
\mathbb{S}^3$ and 
$\mathbb{S}^3 \simeq SU(2)$, 
we can identify 
\[
{\cal M} \simeq \mathbb{S}^2
\]
\[
{\cal H} \simeq U(1)
\]
\[
{\cal G} \simeq SU(2)
\]
so that our starting point is the $SU(2)$ Maurer-Cartan equation
on $\mathbb{S}^2$.

As in Section 2, we first relate the (almost) 
complex structure on $\mathbb{S}^2 \subset \mathbb{R}^3$ to the
vector cross product on $\mathbb{R}^3$. For this we consider
a point $x\in \mathbb{S}^2$, and for any non-vanishing 
$y \in T_x(\mathbb{S}^2)$ so that
\[
<x,y>=0
\] 
we define 
\be
\label{Jvector}
{\mathbb{J}}_x(y) = x \times y
\ee
(Notice that $\mathbb{J}_x (x) = 0$.) Then 
\be
\label{Jcs2}
\mathbb{J}_x^2(y) = x\times(x\times y) = x(x,y)-(x,x)y=-y
\ee
Thus $\mathbb{J}_x^2=-1$ and it defines an almost complex structure.

The matrix realization of $\mathbb{J}_x$ is 
\be
\label{Jmatrix3}
\mathbb{J}_x = \left(
\begin{array}{ccc}
0 & -x_3 & x_2 \\
x_3 & 0 & -x_1 \\
-x_2 & x_1 & 0 \\
\end{array}\right)
\ee
Since $|x|=1$ this clearly satisfies
\be
\label{J1m3}
\left(\mathbb{J}_x^2\right)_{ij} = - \delta_{ij} + x_ix_j
\ee
in full parallel with the construction in Section 2.

Consider now the Maurer-Cartan form for ${\cal G} 
= SU(2)$,
\begin{equation}
A \ = \ g^{-1} d g \ = \ 
\left(
\begin{array}{cc}
\kappa & \vartheta \\
- \bar\vartheta & -\kappa\\
\end{array}\right)
\end{equation}
where we have used the holomorphic polarization
to represent $\vartheta$ in a complex basis. 
The Maurer-Cartan equations 
(\ref{genmc1}), (\ref{genmc2}) become
\begin{equation}
\label{mcesu2}
F \ = \ d\kappa \ = \ \vartheta \wedge\bar\vartheta
\end{equation}
\begin{equation}
(d + 2 \kappa )\wedge \vartheta \ = \ 0 
\la{FU2}
\end{equation}
where the (almost) complex structure is manifest:
Comparing with (\ref{hitc1}) it is obvious that we can identify
$\Omega = \vartheta$ and (\ref{FU2}) clearly implies (\ref{hitc2}).
The condition (\ref{hitc3}) is also satisfied since 
$F$ in (\ref{mcesu2}) is nonvanishing, as its integral over
${\cal M} \simeq \mathbb{S}^2$ is the first Chern class of
the U(1) bundle which is nontrivial as it relates to the area
two-form on the base $\mathbb{S}^2$
\begin{equation}
{\tt Ch}_1(F) \ = \ \frac{1}{2\pi} \int F \ = \ \frac{1}{2\pi}
\int \vartheta \wedge \bar\vartheta
\la{ch1}
\end{equation}
The nontriviality of this first Chern class can also be seen
explicitely, by considering the following two 
parametrizations of the group manifold ${\cal G} 
\simeq SU(2)$: We describe this manifold 
either by using the coordinates
\begin{equation}
\label{gsu2}
g = \left(
\begin{array}{cc}
\alpha & \beta \\
-\bar\beta & \bar\alpha\\
\end{array}\right) \in SU(2)
\end{equation}
where $|\alpha|^2+|\beta|^2=1$, or by employing the 
angular parametrization
\begin{equation}
\alpha = \cos\frac{\theta}{2}\ e^{ \frac{i}{2} (\psi+\phi)} 
\equiv \cos\frac{\theta}{2}\ e^{i\psi_+}
\la{angle1}
\end{equation}
\begin{equation}
\beta = \sin\frac{\theta}{2}\ e^{ \frac{i}{2} (\psi-\phi) } 
\equiv \sin\frac{\theta}{2}\ e^{i\psi_-}
\la{angle2}
\end{equation}
Explicitely, we find for the Maurer-Cartan form
\ba
\label{mcforms}
g^{-1}dg = \left(
\begin{array}{cc}
\kappa & \vartheta \\
-\bar\vartheta & -\kappa\\
\end{array}\right) = \left(
\begin{array}{cc}
\bar\alpha d\alpha+\beta d\bar\beta & \bar\alpha 
d\beta-\beta d\bar\alpha \\
\bar\beta d\alpha-\alpha d\bar\beta & \bar\beta 
d\beta+\alpha d\bar\alpha\\
\end{array}\right)=
\\ \\
= \left(
\begin{array}{cc}
{i\over 2}(\cos\theta d\psi+d\phi) & \2 e^{-i\phi}
(d\theta+i\sin\theta d\psi) \\
-\2 e^{i\phi}(d\theta-i\sin\theta d\psi) & -{i\over 
2}(\cos\theta d\psi+d\phi)\\
\end{array}\right)
\ea
It is a familiar result that (\ref{angle1}), (\ref{angle2})
describe two hemispheres of $\mathbb{S}^2$ which are 
parameterized by angles $(\theta,\psi_\pm)$ and 
$\psi_+-\psi_-=\phi$ that is we have the Hopf 
bundle, or Dirac monopole, with unit charge. Indeed,
in the present case the equations (\ref{genmc1}),
(\ref{genmc2}), (\ref{mcesu2}), (\ref{FU2}) have a 
simple physical interpretation as the familiar 
relation between the vortex condensate $\vartheta$ 
in the background of the abelian gauge field 
$\kappa$, and the ensuing magnetic flux; see also
\cite{fad}.

We note that the Maurer-Cartan forms are complex-valued
one forms on $T^*\mathbb{S}^2$ with coordinates $(\theta,
\psi)$,
\[
\vartheta \ = \ \2 e^{-i\phi}(d\theta+i\sin\theta d\psi)
\]
\[
\bar\vartheta \ = \ \2 e^{i\phi}(d\theta-i\sin\theta d\psi)
\]
and 
\[
\vartheta \wedge \bar\vartheta \ = \ 
-\frac{i}{2}\sin \theta d\theta \wedge d\psi
\]
is the volume two-form on $\mathbb{S}^2$, and the integral
(\ref{ch1}) is clearly nontrivial. In terms of
the cartesian coordinates on $\mathbb{R}^3$ we then have
\[
e^{i\phi} \vartheta \ = \ idx_1+\sum_{a=1,2} 
J_{1\alpha}dx_a
\]
where
\[
J ={\mathbb{J}_x}_{ |_{x_3=\pm \sqrt{1- x_1^2-x_2^2} }}
\]

\section{ $\mathbb{S}^4$ And Instantons}

The four dimensional sphere $\mathbb{S}^4$ is a well
known example of a manifold with {\em no} almost complex structure. 
But there does not seem to be any easy, direct way to see this. 
Instead one exploits the properties of characteristic classes: 
The tangent bundle $T\mathbb{S}^4$ is nontrivial, with a 
nontrivial Euler class. Consequently for an almost complex 
structure the second Chern class ${\tt Ch}_2$ of the complex 
tangent bundle $T\mathbb{S}^4$ should be nontrivial. This Chern class
is proportional to the first Pontryagin class of the underlying
real tangent bundle. But this vanishes since $\mathbb{S}^4$ is a 
hypersurface in $\mathbb{R}^5$ with a trivial normal bundle. 
As a consequence $T\mathbb{S}^4$ cannot be a complex bundle, 
{\it i.e.} $\mathbb{S}^4$ can not have any almost complex 
structure.

We now describe how this conclusion emerges from 
the present formalism. Using the fact that
\[
SO(D+1) / SO(D) \simeq \mathbb{S}^D
\]
we select
\[
{\cal M} \ \simeq \ \mathbb{S}^4
\]
\[
{\cal H} \ \simeq \ SO(4)
\]
\[
{\cal G} \ \simeq \ SO(5)
\] 
With $\gamma_i$ ($i=1,\ldots,5$) the five dimensional 
Euclidean Dirac matrices
\[
\{ \gamma_i , \gamma_j \}_+\ = \ 2 \delta_{ij}
\]
the Lie algebra of $SO(5)$ is represented by the antisymmetric
matrices
\[
M_{ij} \ = \ \frac{1}{4i} [ \gamma_i , \gamma_j] \ \ \ \ \ (i>j)
\]
Explicitly, 
\begin{equation}
[M_{ij} , M_{kl} ] \ = \ f^{(ij)(kl)(mn)} M_{mn}
\ = \ -i ( \delta_{jk}M_{il} - \delta_{ik}
M_{jl} -\delta_{jl} M_{ik} + \delta_{il} M_{jk} )
\la{so5}
\end{equation}
The $\mathfrak{h} \simeq so(4)$ subalgebra is generated 
by $M_{ij}$ for $i,j =1,2,3,4$. The $so(5)$ generators $M_{ij}$ 
with either $i=5$ (or $j=5$) determine the embedding $\mathfrak{m} \simeq 
T\mathbb{S}^4 \subset so(5)$. This is a symmetric embedding since 
the $C^{stu}$ in (\ref{lie3}) vanish. This suggests that 
if there exists a holomorphic polarization 
on $T^* \mathbb{S}^4$ which allows us to write the present
Maurer-Cartan equation (\ref{genmc2}) in a manifestly complex form, 
the ensuing almost complex structure on $\mathbb{S}^4$ could 
be integrable. But as we see below, the absence 
of any almost complex structure on $\mathbb{S}^4$ manifests 
itself as an obstruction for representing the equation 
(\ref{genmc2}) in a holomorphic polarization.  

We now proceed to inspect the present $SO(5) \sim \mathbb{S}^4
\times SO(4)$ version of the Maurer-Cartan equations.
Explicitly, the Maurer-Cartan equations 
\[
d\rho^{m} \ = \ f^{(ij)(kl)(5m)} A^{ij} \wedge A^{kl}
\]
have the form
\begin{equation}
d\rho^{1} \ = \  - i \rho^2 \wedge \kappa^{21} - i \rho^3 
\wedge \kappa^{31} - i \rho^4 \wedge \kappa^{41}
\la{s41}
\end{equation}
\begin{equation}
d\rho^2 \ = \  - i \rho^1 \wedge \kappa^{21} + i \rho^3 
\wedge \kappa^{32} + i \rho^4 \wedge \kappa^{42} 
\la{s42}
\end{equation}
\begin{equation}
d\rho^3 \ = \  - i \rho^4 \wedge \kappa^{43} + i \rho^1 
\wedge \kappa^{31} + i \rho^2 \wedge \kappa^{32} 
\la{s43}
\end{equation}
\begin{equation}
 d\rho^4 \ = \  - i \rho^3 \wedge \kappa^{43} - i \rho^2 
\wedge \kappa^{42} - i \rho^1 \wedge \kappa^{41} 
\la{s44}
\end{equation}
It is a relatively 
straightforward exercise in linear algebra to show that 
it is {\it impossible } to represent the $d\rho^m$ in any
polarization which allows these equations
to be written in a manifestly complex form.  
For example, if we identify
\[
\theta^{12} \ = \ \rho^1 + i \rho^2
\]
\[
\theta^{34} \ = \ \rho^3 + i \rho^4
\]
we find
\begin{equation}
d\theta^{12} - \theta^{12} \wedge \kappa^{21} \ = \ 
- \frac{1}{2}
( \theta^{34} + \bar\theta^{34}) \wedge ( \kappa^{32} - i
\kappa^{31} ) \ + \ \frac{i}{2} ( \theta^{34} - \bar 
\theta^{34}) \wedge ( \kappa^{42} + i \kappa^{41} )
\la{fail}
\end{equation}
where the {\em r.h.s.} is clearly inconsistent with a
holomorphic polarization.  The same conclusion persists if
following \cite{chern2} we substitute in (\ref{s41})-(\ref{s44})
an arbitrary linear combination of the form
\[
\psi^a \ = \ {p^a}_ b \theta^b + {q^a}_ b \bar\theta^b
\]
where ${p^a}_b$ and ${q^a}_b$ are {\it a priori}
unknown functions. Consequently the fact that
$\mathbb{S}^4$ does not admit any almost complex structure
manifests itself in the fact that the Maurer-Cartan
equations do not allow for a complex decomposition.

From (\ref{genmc1}) we can compute the second Chern 
number of the $SO(4)$ principal bundle over $\mathbb{S}^4$. Since 
\[
f^{(ij)(k5)(m5)} = i
\] 
we have 
\[
F_\kappa^{(ij)} \ = \ -2i \sum_{i<j} \rho^i \wedge \rho^j
\]
and thus the second Chern number for the $SO(4)$ connection
$\kappa$ is computed by the volume four-form on $\mathbb{S}^4$,
\begin{equation}
{\tt Ch}_2 (\kappa) \ =  \ - \frac{1}{8\pi^2} \int {\rm Tr}\ F^2_\kappa
= - \frac{1}{24\pi^2} \int \rho^1 \wedge \rho^2 
\wedge \rho^3 \wedge \rho^4
\la{ch2}
\end{equation}

We observe that the structure of the equations (\ref{s41})-(\ref{s44}) 
coincides with the structure of the equations that 
describe the familiar Yang-Mills instanton bundle 
with base $\mathbb{S}^4$, fibre $SU(2)\simeq \mathbb{S}^3$ and the total
space $\mathbb{S}^7$. 
Indeed, this is expected since the construction of the 
instanton bundle employs
the natural $O(4) \simeq SU(2) \times SU(2)$ connection 
$(\omega_{0i} , \omega_{ij})$ with $i,j =1,2,3$, on the
sphere $\mathbb{S}^4$. 
To show this, we recall the de Sitter
metric on $\mathbb{S}^4$ with unit radius (see e.g. \cite{eguchi}),
\[
ds^2 \ = \ {dr^2+r^2(\varpi_1^2+\varpi_2^2+\varpi_3^2)
\over (1+r^2)^2} \ = \ \sum\limits_{a=0}^3 (e^a)^2
\]
where $\varpi_i$ for $i=1,2,3$ and $e^a$ for $a=0,1,2,3$
are defined by
\begin{equation}
(1+r^2)\left(
\begin{array}{c}
e^0 \\
e^1 \\
e^2 \\
e^3 \\
\end{array}\right)
\ = \
\left(
\begin{array}{c}
dr \\
r\varpi_1 \\
r\varpi_2 \\
r\varpi_3
\\
\end{array}\right)
\ = \
\frac{1}{r} 
\left( 
\begin{array}{cccc}
x & y & z & t \\
-t & -z & y & x \\
z & -t & -x & y \\
-y & x & -t & z \\
\end{array}
\right)
\left(
\begin{array}{c}
dx \\ dy \\ dz \\ dt \\
\end{array} \right)
\la{egh}
\end{equation}
The $e^a$ define the $\mathbb{S}^4$
vierbein, and if we introduce the natural connection 
\[
{\omega^1}_0 = {\omega^2}_3 \ = \ \varpi_1 \ , \ \ \ \ \ \ \ 
{\omega^2}_0 = {\omega^3}_1 \ = \ \varpi_2\ , \ \ \ \ \ \ \ 
{\omega^3}_0 = {\omega^1}_2 \ = \ \varpi_3 
\]
the component form of the ensuing first Cartan structure equation 
(for vanishing torsion $T^a$) 
\begin{equation}
de^a + {\omega^a}_{b} \wedge e^b \ = \ T^a \ = \ 0
\la{cart}
\end{equation}
coincides with the functional 
form of (\ref{s41})-(\ref{s44}). Explicitly,
\[
de^1 \ = \ e^2 \wedge {\omega^1}_ 2 + e^3 \wedge {\omega^2}_ 3
+ e^0 \wedge {\omega^1}_ 0 \ \ \ \ \ (cyclic)
\]
which clearly reveals that we can identify
\begin{equation}
e^i \sim \rho^i \ \ \ \ \ \& \ \ \ \ \ \ e^0 \ \sim \ \rho^4
\la{inid}
\end{equation}
The absence of an almost complex structure on $\mathbb{S}^4$ is 
then synonymous to the fact, that the $\mathbb{S}^4$ Cartan
structure equation (\ref{cart}) is not consistent with any
holomorphic polarization. 

Indeed, this lack of almost complex structures
on $\mathbb{S}^4$ is encoded in the explicit form of the
BPST instanton solution: The self-dual 
and anti-self-dual combinations 
of the connection one-form ${\omega}_{ab}$
\[
A^a_\pm \ = \  \mp \omega_{0a} - \epsilon_{abc} \omega_{bc}
\]
exactly coincide with the components of the 
Yang-Mills connection one-form that describes 
the $SU(2)$ BPST single-instanton solution. 
Furthermore, using the relation between (\ref{cart}) 
and (\ref{s41})-(\ref{s44}), we get from (\ref{ch2}) 
the second Chern number of the BPST instanton in terms
of the volume four-form on the base manifold $\mathbb{S}^4$,
\[
{\tt Ch}_2 (A_\pm) \ \propto \ \int e^0 \wedge e^1 \wedge e^2 \wedge
e^3
\]
This reproduces (\ref{ch2}) when we recall the identification
(\ref{inid}).

\section{Complex Structure of $\mathbb{CP}^2$  }

Complex projective spaces $\mathbb{CP}^n$ with $n\geq 1$ are 
familiar examples of manifolds with nontrivial complex structures
that can be described explicitely, in terms of complex coordinates. 
Here we explain how these complex structures can also 
be understood from
the point of view of the present formalism. For this we consider 
in detail the simplest nontrivial case of $\mathbb{CP}^2$.

There are several ways to introduce complex coordinates
on the manifold $\mathbb{CP}^2$. A particularly instructive one
\cite{last} is obtained by embedding 
\[
(Z_0,Z_1, Z_2) \ \mapsto \ \frac{(Z_2 \bar Z_0 , \ Z_0 \bar Z_1 , \
Z_1 \bar Z_2 , \ |Z_1|^2 - |Z_2|^2)}{|Z_0|^2 + |Z_ 1|^2 + 
|Z_ 2|^2}
\]
of $\mathbb{CP}^2$ into 
$\mathbb{R}^7$, which is considered as a subspace of 
$ \mathbb{C}^4$. The ensuing complex coordinates
on $\mathbb{CP}^2$ are then ratios of the homogenenous coordinates
$(Z_0, Z_1, Z_2)$.
In contrast to $\mathbb{S}^4$ where one adds a single infinity point
to $\mathbb{R}^4$, in the case of $\mathbb{CP}^2$ one adds at infinity
the two-dimensional cycle $\mathbb{CP}^1 \simeq 
\mathbb{S}^2$, which gives an additional possibility of
regluing the complex co-ordinates
\footnote{The open set of $\mathbb{CP}^2$ around the "origin" can be described
in terms of the co-ordinates $(z_1=Z_1/Z_0,z_2=Z_2/Z_0)$. 
At $z_2\to\infty$ one should use the complex 
co-ordinates $u=(u_1,u_2)$: $u_1=z_1/z_2=Z_1/Z_2$ 
and $u_2=1/z_2=Z_0/Z_2$. Contrarily, at 
$z_1\to\infty$ one should use instead the co-ordinates $v=(v_1,v_2)$, 
where $v_1=1/z_1=Z_0/Z_1$ and $v_2=z_2/z_1=Z_2/Z_1$. 
In the "symmetric" $\mathbb{CP}^2$ 
case at $z\to\infty$ the whole $\mathbb{CP}^1$ is present, where 
$u$ and $v$ complex co-ordinates van be 
"reglued" into each other, however if we "remove" this 
line (the $\mathbb{S}^4$ 
case) the whole picture becomes singular, since one cannot just put $u=v$.}.

Clearly, this is a generalization of the familiar representation 
of $\mathbb{CP}^1 \simeq \mathbb{S}^2 \subset \mathbb{R}^3$ 
in terms of a stereographic projection.

In order to describe the complex structure 
of $\mathbb{CP}^2$ in terms of a Yang-Mills theory, 
we start from the fact that locally
\[
SU(3)  \sim U(2) \times \mathbb{CP}^2
\] 
so that we have the identifications
\[
{\cal M} \ \simeq \ \mathbb{CP}^2
\]
\[
{\cal H} \ \simeq \ U(2)
\]
\[
{\cal G} \ \simeq \ SU(3)
\] 
We then consider the Maurer-Cartan
equation of the $SU(3)$ Yang-Mills theory. 
In the standard Gell-Mann basis the subalgebra 
$u(2) \sim su(2)\times u(1)$ is 
generated by the matrices $\lambda_a$ for 
$a=1,2,3,8$. Since the $C^{stu}$ in (\ref{lie3}) 
now vanish the embedding of $U(2) \subset SU(3)$ 
is symmetric which suggests that an almost complex 
structure can be integrable.

We introduce the holomorphic polarization
\[
\theta_1 \ = \ \rho_4 + i \rho_5
\]
\[
\theta_2 \ = \ \rho_6 + i \rho_7
\]
This leads to the following version of the 
Maurer-Cartan equation
(\ref{genmc1})
\begin{equation}
F_\kappa^1 + i F_\kappa^2 \ = \ 
- \frac{i}{2} \theta_1 \wedge \bar\theta_2
\la{cp1}
\end{equation}
\begin{equation}
F_\kappa^3 \ = \ - \frac{i}{4} \theta_1 \wedge \bar\theta_1
+ \frac{i}{4} \theta_2 \wedge \bar\theta_2
\la{cp2}
\end{equation}
\begin{equation}
F_\kappa^8 \ = \ i \frac{\sqrt{3}}{4} \theta_1 \wedge
\bar\theta_1 - i \frac{\sqrt{3}}{4} \theta_2 \wedge 
\bar\theta_2
\la{cp3}
\end{equation}
and for (\ref{genmc2}) we get
\begin{equation}
d \theta_1 \ = \ - i \kappa_3 \wedge \theta_1 - 
i \sqrt{3} \kappa_8 \wedge \theta_1
- i(\kappa_1 + i \kappa_2) \wedge \theta_2 
\la{cp4}
\end{equation}
\begin{equation}
d \theta_2 \ = \ i \kappa_3 \wedge \theta_2 - 
i \sqrt{3} \kappa_8 \wedge \theta_2  - i(\kappa_1 - i \kappa_2)
\wedge \theta_1 
\la{cp5}
\end{equation}
Here the almost complex structure of $\mathbb{CP}^2$ is
manifest. Furthermore, comparing with (\ref{hitc2}) 
we conclude that this almost complex structure is indeed
integrable.

We note that the equations (\ref{cp1})-(\ref{cp3}) have the structure
of (\ref{ws1}). The equations (\ref{cp4}), (\ref{cp5}) can also
be combined into the functional form (\ref{ws2}), 
\[
D_\kappa \Phi \ = \ 0
\]
when we define the $U(2)\sim U(1) \times SU(2)$ covariant 
derivative
\[
D_\kappa \ = \ I \otimes d + i I \otimes \sqrt{3}\kappa_8 +
i \sigma_i \otimes \kappa_i 
\]
where the $\sigma_i$ are Pauli matrices, and 
\[
\Phi \ = \ \begin{pmatrix}  \theta_1 \\ \theta_2 
\end{pmatrix}
\]

Finally, the Chern classes of the $U(2)$
Yang-Mills connection $\kappa$ can be computed directly
from the cohomology classes of $\mathbb{CP}^2$, 
from (\ref{cp1})-(\ref{cp3}) we find for the second Chern class
\[
{\tt Ch}_2 \ = \ \frac{1}{8\pi^2}
\int \left( {\rm Tr} F_\kappa^2 - {\rm Tr} F_\kappa {\rm Tr} F_\kappa \right)
\ \propto \ \int \theta_1 \wedge \theta_2 \wedge \bar\theta_1 
\wedge \bar\theta_2
\]

\section{Conclusions }

In conclusion, Chern's attempt to prove his Last Theorem which states 
that $\mathbb{S}^6$ has no complex structure has a natural 
interpretation in terms of Yang-Mills theory. Unfortunately,
the completion of his proof appears to be highly nontrivial but 
to argue to the contrary it seems to us, that an explicit 
realization of a complex structure needs to be constructed.
Here we have proposed how his arguments can be viewed 
in a wider prespective, and considered several examples. In particular,
we have suggested that Chern's approach relates directly to 
the study of physically relevant issues, such as the unified description
of space-time and gauge symmetry and geometrical 
structure of an unbroken vacuum in a spontaneously broken 
gauge theory. One may also hope that this leads to a
better understanding of the
relation between the Yang-Mills theory and Kodaira-Spencer theory
of gravity, relevant for the study of topological strings.
Consequently we expect that further investigations 
of the relations between the geometry of manifolds, complex 
structures, and the Yang-Mills theory along these lines could 
lead to interesting and both physically and mathematically relevant 
observations.

\section{ Acknowledgements}

A.N. is indebted to Shiing-Shen Chern for describing
his Last Theorem and for providing a copy of the preprint.
Shiing-Shen Chern passed away only a few weeks after he completed
the proof, and we dedicate this paper to his memory.
We are also grateful to C.~Bai, V.~Dolotin, G.~Etesi, V.~Fock, M.~Gunaydin, 
S.~Khoroshkin, J.~Minahan, N.~Nekrasov,
P.~Pushkar and V.~Rubtsov for useful discussions and communications.
A.M. thanks Department of Theoretical Physics at Uppsala University
for the warm hospitality during initial stage of this work, and
A.N. thanks Mo Lin Ge for warm hospitality at Nankai Institute
of Mathematics, and Tohru Eguchi for warm hospitality at
Tokyo University. This work has been supported by a VR Grant, 
by a STINT Thunberg scholarship, and by a STINT Institutional Grant 
IG2004-2 025. The work of A.M. was also
partially supported by RFBR grant 02-02-16496, the grant for 
support of Scientific 
Schools 1578.2003.2, and the Russian Science Support Foundation.

\vfill\eject

\end{document}